\documentclass[12pt]{article}

\newtheorem{theorem}{Theorem}{\bf }{\it }
{\bf }{\it }

\newcommand{\np}{\mbox{NP}}
\newcommand{\p}{\mbox{P}}

\title{Formalization of the class of problems\\ solvable 
by a nondeterministic Turing machine}
\author{Anatoly D. Plotnikov\thanks{This paper is published in
the journal ``Cybernetics and Systems Analysis.'' Vol. 33, No. 5, 1997, 
pp. 635--640. Translated by Plenum Publishing Corporation, NY.}
}
\date{}

\begin{document}
\maketitle

\begin{abstract}
The objective of this article is to formalize the definition of \np\ problems.

We construct a mathematical model of discrete problems as independence systems 
with weighted elements. We introduce two auxiliary sets that characterize the 
solution of the problem: the adjoint set, which contains the elements from the
original set none of which can be adjoined to the already chosen solution 
elements; and the residual set, in which every element can be adjoined to 
previously chosen solution elements.

In a problem without lookahead, every adjoint set can be generated by the 
solution algorithm effectively, in polynomial time. 

The main result of the study is the assertion that the \np\ class is identical
with the class of problems without lookahead. Hence it follows that if we fail 
to find an effective (polynomial-time) solution algorithm for a given problem,
then we need to look for an alternative formulation of the problem in set of
problems without lookahead.
\end{abstract}

\section{Introduction}

Solvability is the key problem in the theory of solution of discrete problems 
\cite{1} -- \cite{7}. The input data and the solution result for any discrete 
problem are usually finite, and generally discrete problems do not suffer from 
the classical difficulty of total nonexistence of a solution algorithm. Many 
discrete problems have a trivial solution algorithm, which involves exhaustive 
enumeration of the elements of the solution. In practice, however, the trivial 
algorithm is inapplicable, because its computational complexity is too high 
even for a relatively small number of solution elements. Thus, if every 
discrete problem is interpreted as a problem of constructing a subset that 
satisfies given constraints among the elements of some initial $n$-set, then 
the trivial algorithm in general requires inspecting all $2^{n}$ subsets, which 
is obviously intractable. In such cases, we say that the trivial algorithm 
runs in exponential time, or has exponential complexity.

In the context of solvability of discrete problems we usually discuss the 
possibility of developing an algorithm that generates a solution in a time 
essentially shorter than the running time of the trivial algorithm. A discrete 
problem is regarded as effectively solvable if the running time of the solution 
algorithm is polynomial in the size of the problem. Such algorithms are called 
polynomial-time algorithms. Here and in what follows, the size of a discrete 
problem is the number n of elements in the input set.

The difficulties that arise in the process of development of solution 
algorithms for various discrete problems have led to the identification of a 
class of problems for which it is expedient to look for effective or 
polynomial-time algorithms. First, all problems for which no solution algorithm 
exists (e.g., solvability of polynomial equations in integers) or for which the 
number of solutions depends exponentially on the size of the problem (e.g., 
finding all $2^{n-2}$ covering trees of an $n$-graph) have been excluded 
\cite{2, 3}. Among the set of discrete problems for which the number of 
solution elements (the length of the solution) is a polynomial function of 
problem size we focus on problems that are solvable by a nondeterministic 
Turing machine (NTM) in polynomial time. Discrete problems satisfying these 
constraints form the class \np.

Effective (polynomial-time) algorithms are available for solving some \np\ 
problems, and we accordingly identify a subclass $\p\subset \np$ of problems 
with polynomial-time algorithms. For many practically important \np\ problems, 
however, attempts to find effective solution algorithms have failed.

The issue of discrete-problem solvability thus involves the relationship 
between the class \np\ and its subclass \p. Some authors maintain that a strict 
inclusion applies, i.e., $P\subset \np$ and $P\not= \np$, while others claim 
that \p = \np. This disagreement among mathematicians is primarily due to 
ambiguously defined notions of NTM operation. The objective of this article 
is to formalize the definition of \np\ problems.

We consider the solution of an individual enumerative \np\ problem. By analyzing 
the operation of the NTM in the process of solving the problem, we establish 
that different interpretations of NTM operation lead to an ambiguous 
description of the process. In one interpretation, the NTM constructs the next 
intermediate solution using only the previously generated computation results; 
the other interpretation ignores this important feature.

We thus establish that unacceptably large enumeration during problem solving 
arises only when the NTM chooses the next solution element by inspecting an 
exponential number of all (final or support) solutions. The set of discrete 
problems is thus partitioned into two disjoint sets: problems for which the 
next solution element can be chosen without inspecting all the support 
solutions; and problems for which such choice is impossible. Problems of the 
first class are called problems without lookahead, while problems of the second 
class are called inherently exponential.

We construct a mathematical model of discrete problems as independence systems 
with weighted elements. We introduce two auxiliary sets that characterize the 
solution of the problem: the adjoint set, which contains the elements from the
original set none of which can be adjoined to the already chosen solution 
elements; and the residual set, in which every element can be adjoined to 
previously chosen solution elements.

In a no-lookahead problem, every adjoint set can be generated by the solution 
algorithm effectively, in polynomial time. 

The main result of the study is the assertion that the \np\ class is identical 
with the class of problems without lookahead. Hence it follows that if we fail 
to find an effective (polynomial-time) solution algorithm for a given problem, 
then we need to look for an alternative formulation of the problem in the set 
of no-lookahead problems.

\section{Example of a discrete problem}

Consider the acyclic digraph shown in Fig. \ref{11} (a) (in all figures, the 
arcs are directed from bottom up). The transitive closure graph of the acyclic 
digraph is obviously a graph of a strict partial ordering. Any algorithm that 
constructs the maximum matching in a bipartite graph (see, e.g., \cite{6}) can 
be applied to partition the nodes of this graph into a minimum number of chains 
(a so-called minimum chain partition, MCP). One of these MCPs contains the 
nodes and the arcs of the transitive closure graph that are shown by thick 
lines in Fig. \ref{11} (a).

\begin{figure}[htbp]
\centering
\unitlength 1.00mm
\linethickness{0.1pt}
\begin{picture}(94.94,94.00)
\put(10.00,10.00){\line(0,1){30.00}}
\put(10.00,40.00){\line(2,-3){20.00}}
\put(10.15,40.13){\line(2,-3){20.00}}
\put(10.30,40.26){\line(2,-3){20.00}}
\put(9.85,39.87){\line(2,-3){20.00}}
\put(9.70,39.74){\line(2,-3){20.00}}
\put(30.00,10.00){\line(0,1){30.00}}
\put(30.00,40.00){\line(-2,-1){20.00}}
\put(30.15,39.87){\line(-2,-1){20.00}}
\put(30.30,39.74){\line(-2,-1){20.00}}
\put(29.85,40.13){\line(-2,-1){20.00}}
\put(29.70,40.26){\line(-2,-1){20.00}}
\put(10.00,20.00){\line(1,1){20.00}}
\put(10.00,10.00){\line(2,3){20.00}}
\put(30.00,30.00){\line(1,1){10.00}}
\put(30.15,29.87){\line(1,1){10.00}}%
\put(30.30,29.74){\line(1,1){10.00}}%
\put(29.85,30.13){\line(1,1){10.00}}%
\put(29.70,30.26){\line(1,1){10.00}}%
\put(40.00,40.00){\line(0,-1){30.00}}
\put(40.00,10.00){\line(-1,1){10.00}}
\put(40.15,10.13){\line(-1,1){10.00}}
\put(40.30,10.26){\line(-1,1){10.00}}
\put(39.85,9.87){\line(-1,1){10.00}}
\put(39.70,9.74){\line(-1,1){10.00}}
\put(30.00,20.00){\line(1,2){10.00}}
\put(40.00,10.00){\line(-1,2){10.00}}
\put(10.00,20.00){\oval(15.00,20.00)[l]}
\put(30.00,20.00){\oval(15.00,20.00)[l]}
\put(30.00,30.00){\oval(12.00,20.00)[l]}
\put(30.00,25.00){\oval(10.00,30.00)[r]}
\put(10.00,10.00){\circle*{3.00}}
\put(30.00,10.00){\circle*{3.00}}
\put(40.00,10.00){\circle*{3.00}}
\put(10.00,20.00){\circle*{3.00}}
\put(30.00,20.00){\circle*{3.00}}
\put(10.00,30.00){\circle*{3.00}}
\put(30.00,30.00){\circle*{3.00}}
\put(10.00,40.00){\circle*{3.00}}
\put(30.00,40.00){\circle*{3.00}}
\put(40.00,40.00){\circle*{3.00}}
\put(12.50,6.00){\makebox(0,0)[cc]{$x_{1}$}}
\put(32.50,6.00){\makebox(0,0)[cc]{$x_{2}$}}
\put(42.50,6.00){\makebox(0,0)[cc]{$x_{3}$}}
\put(6.00,17.00){\makebox(0,0)[cc]{$x_{4}$}}
\put(27.50,17.50){\makebox(0,0)[cc]{$x_{5}$}}
\put(5.11,33.33){\makebox(0,0)[cc]{$x_{6}$}}
\put(27.00,26.50){\makebox(0,0)[cc]{$x_{7}$}}
\put(12.50,44.00){\makebox(0,0)[cc]{$x_{8}$}}
\put(32.50,44.00){\makebox(0,0)[cc]{$x_{9}$}}
\put(42.50,44.00){\makebox(0,0)[cc]{$x_{10}$}}
\put(10.00,60.00){\line(0,1){30.00}}
\put(10.00,90.00){\line(2,-3){20.00}}
\put(30.00,60.00){\line(0,1){30.00}}
\put(30.00,90.00){\line(-2,-1){20.00}}
\put(10.00,70.00){\line(1,1){20.00}}
\put(10.00,60.00){\line(2,3){20.00}}
\put(30.00,80.00){\line(1,1){10.00}}
\put(40.00,90.00){\line(0,-1){30.00}}
\put(40.00,60.00){\line(-1,1){10.00}}
\put(30.00,70.00){\line(1,2){10.00}}
\put(40.00,60.00){\line(-1,2){10.00}}
\put(10.00,70.00){\oval(15.00,20.00)[l]}
\put(30.00,70.00){\oval(15.00,20.00)[l]}
\put(30.00,80.00){\oval(12.00,20.00)[l]}
\put(30.00,75.00){\oval(10.00,30.00)[r]}
\put(10.00,60.00){\circle*{3.00}}
\put(30.00,60.00){\circle*{3.00}}
\put(40.00,60.00){\circle*{3.00}}
\put(10.00,70.00){\circle*{3.00}}
\put(30.00,70.00){\circle*{3.00}}
\put(10.00,80.00){\circle*{3.00}}
\put(30.00,80.00){\circle*{3.00}}
\put(10.00,90.00){\circle*{3.00}}
\put(30.00,90.00){\circle*{3.00}}
\put(40.00,90.00){\circle*{3.00}}
\put(12.50,56.00){\makebox(0,0)[cc]{$x_{1}$}}
\put(32.50,56.00){\makebox(0,0)[cc]{$x_{2}$}}
\put(42.50,56.00){\makebox(0,0)[cc]{$x_{3}$}}
\put(6.00,67.00){\makebox(0,0)[cc]{$x_{4}$}}
\put(27.50,67.50){\makebox(0,0)[cc]{$x_{5}$}}
\put(5.11,83.33){\makebox(0,0)[cc]{$x_{6}$}}
\put(27.00,76.50){\makebox(0,0)[cc]{$x_{7}$}}
\put(12.50,94.00){\makebox(0,0)[cc]{$x_{8}$}}
\put(32.50,94.00){\makebox(0,0)[cc]{$x_{9}$}}
\put(42.50,94.00){\makebox(0,0)[cc]{$x_{10}$}}
\put(62.44,70.00){\line(2,-3){20.00}}
\put(82.44,70.00){\line(-2,-1){20.00}}
\put(82.44,60.00){\line(1,1){10.00}}
\put(92.44,70.00){\line(0,-1){30.00}}
\put(92.44,40.00){\line(-1,1){10.00}}
\put(82.44,40.00){\circle*{3.00}}
\put(92.44,40.00){\circle*{3.00}}
\put(82.44,50.00){\circle*{3.00}}
\put(62.44,60.00){\circle*{3.00}}
\put(82.44,60.00){\circle*{3.00}}
\put(62.44,70.00){\circle*{3.00}}
\put(82.44,70.00){\circle*{3.00}}
\put(92.44,70.00){\circle*{3.00}}
\put(84.94,36.00){\makebox(0,0)[cc]{$x_{2}$}}
\put(94.94,36.00){\makebox(0,0)[cc]{$x_{3}$}}
\put(86.83,53.06){\makebox(0,0)[cc]{$x_{5}$}}
\put(64.00,55.77){\makebox(0,0)[cc]{$x_{6}$}}
\put(79.44,56.50){\makebox(0,0)[cc]{$x_{7}$}}
\put(64.94,74.00){\makebox(0,0)[cc]{$x_{8}$}}
\put(84.94,74.00){\makebox(0,0)[cc]{$x_{9}$}}
\put(94.94,74.00){\makebox(0,0)[cc]{$x_{10}$}}
\put(77.78,30.67){\makebox(0,0)[cc]{Alternating cycle}}
\put(24.67,51.11){\makebox(0,0)[cc]{(a)}}
\put(77.56,25.11){\makebox(0,0)[cc]{(b)}}
\put(24.67,2.67){\makebox(0,0)[cc]{(c)}}
\linethickness{3.0pt}
\put(62.67,60.00){\line(0,1){10.00}}
\put(82.67,39.78){\line(0,1){10.44}}
\put(82.67,59.56){\line(0,1){10.22}}
\put(10.00,60.00){\line(0,1){30.00}}
\put(30.00,60.00){\line(0,1){30.00}}
\put(40.00,90.00){\line(0,-1){30.00}}
\put(92.44,70.00){\line(0,-1){30.00}}
\put(10.00,10.00){\line(0,1){20.00}}
\put(30.00,20.00){\line(0,1){10.00}}
\end{picture}
\caption{}
\label{11}
\end{figure}

It is easy to see that the first chain in the MCP starting from the node 
$x_{1}$ contains two pairs of independent digraph nodes: $x_{1}$, $x_{8}$ 
and $x_{4}$, $x_{8}$. In general, the transitive closure graph can have several 
different MCPs, and the transition from one MCP to another is possible if we 
find an alternating cycle or an alternating chain.

Suppose that in the transitive closure graph for a given MCP it is required to 
find an alternating chain or an alternating cycle that takes us from the 
current MCP to another MCP such that none of the chains contains independent 
nodes of the original graph.

Is this an \np\ problem? The solution of this problem -- an alternating cycle -- 
obviously contains a number of elements that depends linearly on the number of 
nodes of the acyclic digraph. The problem is thus \np\ if it is solvable by NTM 
in polynomial time.

According to one interpretation, the NTM operates in two stages \cite{3, 4, 6}: 
first the machine ``guesses'' some sequence of solution elements, and then it 
decides in polynomial time that the guessed sequence is a solution of the given 
problem. It is thus assumed that the guessing stage and the verification stage 
are both executed by the NTM in polynomial time. A key point in this 
interpretation is the feasibility of deciding in polynomial time that the 
solution is correct.

The correctness of the presented solution -- an alternating cycle -- obviously 
can be checked in polynomial time for our problem. Therefore, according to 
this interpretation, the problem is \np.

Yet there is also an alternative interpretation of NTM operation \cite{1, 5, 7}. 
In \cite{5}, the operation of an NTM is illustrated by the problem of finding 
a correct $k$-coloring of some $n$-graph. There are $k^{n}$ different ways to 
paint the nodes of an $n$-graph in $k$ colors. It is required to decide if at 
least one of these colorings is a correct coloring. To this end it is obviously 
sufficient to examine all the edges of the colored graph, and if the end points 
of each edge are painted in different colors, then the coloring is correct.

The number of edges in a graph is of order $O(n^{2})$. According to this NTM 
interpretation, we need to check simultaneously all $k^{n}$ colorings, and the 
entire checking procedure is a linear function of the length of the input 
data -- the number of elements in the adjacency matrix of the graph.

Under this interpretation, the main distinction between the operation of the 
NTM and the operation of a deterministic Turing machine (DTM) is that the NTM 
checks concurrently the correctness of all alternatives. Curiously, however, 
some authors (see, e.g., \cite{1}) use both interpretations of NTM operation 
simultaneously.

Note that if we adopt the second interpretation of NTM operation, then the NTM 
goes from one state to the next only on the basis of previously generated 
computation results. This is also confirmed by simulating the operation of an 
NTM in a DTM with exhaustive enumeration of all computations \cite{1}. We know 
that in each computation step the DTM goes from one state to another (and 
writes appropriate records on the output tape) only on the basis of previously 
generated (intermediate) results.

Let us consider from this point of view the construction of the alternating 
cycle shown in Fig. \ref{11} (b). (Figure \ref{11} (c) shows the MCP generated 
by this cycle.) Figure \ref{22} (a) shows a part of this cycle, and Fig. 
\ref{22} (b) shows the digraph with the elements of the partition of its 
transitive closure graph into chains generated by an intermediate computation 
result.

\begin{figure}[htbp]
\centering
\unitlength 1.00mm
\linethickness{0.1pt}
\begin{picture}(82.06,44.44)
\put(49.56,10.00){\line(0,1){30.00}}
\put(49.56,40.00){\line(2,-3){20.00}}
\put(49.71,40.12){\line(2,-3){20.00}}%
\put(49.86,40.24){\line(2,-3){20.00}}%
\put(49.41,39.88){\line(2,-3){20.00}}%
\put(49.26,39.76){\line(2,-3){20.00}}%
\put(69.56,10.00){\line(0,1){30.00}}
\put(69.56,40.00){\line(-2,-1){20.00}}
\put(49.56,20.00){\line(1,1){20.00}}
\put(49.56,10.00){\line(2,3){20.00}}
\put(69.56,30.00){\line(1,1){10.00}}
\put(79.56,40.00){\line(0,-1){30.00}}
\put(79.56,10.00){\line(-1,1){10.00}}
\put(69.56,20.00){\line(1,2){10.00}}
\put(79.56,10.00){\line(-1,2){10.00}}
\put(49.56,20.00){\oval(15.00,20.00)[l]}
\put(69.56,20.00){\oval(15.00,20.00)[l]}
\put(69.56,30.00){\oval(12.00,20.00)[l]}
\put(69.56,25.00){\oval(10.00,30.00)[r]}
\put(49.56,10.00){\circle*{3.00}}
\put(69.56,10.00){\circle*{3.00}}
\put(79.56,10.00){\circle*{3.00}}
\put(49.56,20.00){\circle*{3.00}}
\put(69.56,20.00){\circle*{3.00}}
\put(49.56,30.00){\circle*{3.00}}
\put(69.56,30.00){\circle*{3.00}}
\put(49.56,40.00){\circle*{3.00}}
\put(69.56,40.00){\circle*{3.00}}
\put(79.56,40.00){\circle*{3.00}}
\put(52.06,6.00){\makebox(0,0)[cc]{$x_{1}$}}
\put(72.06,6.00){\makebox(0,0)[cc]{$x_{2}$}}
\put(82.06,6.00){\makebox(0,0)[cc]{$x_{3}$}}
\put(45.56,17.00){\makebox(0,0)[cc]{$x_{4}$}}
\put(67.06,17.50){\makebox(0,0)[cc]{$x_{5}$}}
\put(44.67,33.33){\makebox(0,0)[cc]{$x_{6}$}}
\put(66.56,26.50){\makebox(0,0)[cc]{$x_{7}$}}
\put(52.06,44.00){\makebox(0,0)[cc]{$x_{8}$}}
\put(72.06,44.00){\makebox(0,0)[cc]{$x_{9}$}}
\put(82.06,44.00){\makebox(0,0)[cc]{$x_{10}$}}
\put(9.78,40.44){\line(2,-3){20.00}}
\put(29.78,10.44){\circle*{3.00}}
\put(29.78,20.44){\circle*{3.00}}
\put(9.78,30.44){\circle*{3.00}}
\put(9.78,40.44){\circle*{3.00}}
\put(32.28,6.44){\makebox(0,0)[cc]{$x_{2}$}}
\put(29.94,25.94){\makebox(0,0)[cc]{$x_{5}$}}
\put(10.00,25.77){\makebox(0,0)[cc]{$x_{6}$}}
\put(12.28,44.44){\makebox(0,0)[cc]{$x_{8}$}}
\put(18.89,2.44){\makebox(0,0)[cc]{(a)}}
\put(61.78,2.44){\makebox(0,0)[cc]{(b)}}
\linethickness{3.0pt}
\put(9.78,30.67){\line(0,1){9.56}}
\put(29.56,10.67){\line(0,1){9.56}}
\put(49.56,10.00){\line(0,1){20.00}}
\put(69.56,20.00){\line(0,1){20.00}}
\put(79.56,40.00){\line(0,-1){30.00}}
\end{picture}
\caption{}
\label{22}
\end{figure}

It is easy to see that the next thin arc $(x_{3}, x_{5})$ of the alternating 
cycle cannot be chosen unless we know in advance that the ``thick'' arc 
$(x_{7}, x_{9})$ will subsequently be included in the cycle being constructed. 
Thus, according to the second interpretation of NTM operation, this problem is 
not \np.

We have reached a diametrically opposite conclusion to the previous one. To 
eliminate the contradiction, we need to formalize the class of problems 
solvable by NTM in polynomial time.

The second interpretation of NTM operation is more appealing. Thus, in the 
example of checking the correctness of a graph coloring it is natural to 
assume that in each computation step the NTM decides which of the presented 
colorings are correct, and in the next step the decision about new correct 
colorings is reached by analyzing only the new edge in each ``correct'' option. 
If we adopt the other interpretation, then we have to agree that the NTM has 
an ``instantaneous solver'' that in each step allows the machine to ``look 
ahead'' into the required answer and thus decide which of the intermediate 
options is correct and which is not.

We have previously identified the ``uninteresting'' class of discrete problems 
that are inherently exponential. The problem of finding an alternating chain 
or cycle should be classified as inherently exponential, because in this 
problem we cannot use a partial (intermediate) result to pass to the next 
``correct'' intermediate or final result. In general, this problem requires 
``inspecting'' all final results (the number of which is an exponential function 
of the number of graph nodes).

In the next section we formalize the characteristics of such problems.

\section{Set-theoretical model of discrete problems}

In set-theoretical terms, many (finite) discrete problems involve selecting 
subsets $\pi_{i}$ $(i=\overline {1, m})$ from some $n$-set $R$ that satisfy 
given constraints \cite{8}. A discrete problem is therefore defined as the 
4-tuple $Z = (R. Q, M,f)$, where $R = \{r_{1},\ldots, r_{n}\}$ $(n > 1)$ is 
called the {\it work set} and the {\it feasibility region} $Q$ is a nonempty 
collection of subsets $\pi$ of the set $R$ that satisfy the following 
condition:

$1^{0}$) if $\pi\in Q$ and $\pi_{1}\subset \pi$, then $\pi_{1}\in Q$.

The set $M = \{\mu(r_{1}), \ldots, \mu(r_{n})\}$ is a collection of nonnegative 
integers, and $f$ is a function defined on $Q$. The pair $(R, Q)$ is obviously 
an {\it independence system}.

Each element $\pi$ in $Q$ is called a feasible solution of problem $Z$, and the 
number $\mu(r_{i})\in M$ $(\mu(r_{i}) > 0$ is the weight of the element 
$r_{i}\in R$ $(i = \overline {1, n})$.

In what follows we assume that for every $\pi\in Q$,
\[
f(\pi)=\sum_{\forall r_{i}\in \pi} \mu(r_{i}).
\]

The solution $\pi\in Q$ is called a {\it support solution} if there is no 
$\pi_{1}\in Q$ such that $\pi\subset \pi_{1}$ and $\pi$ is a proper subset of 
the set $\pi_{1}$. Denote by $B\subset Q$ the set of all support solutions of 
problem $Z$.

Problem $Z$ is called {\it nontrivial} if $Card(B) = O(2^{p(n)})$. In other 
words, problem $Z$ is nontrivial if the set of its support solutions contains 
exponentially many elements.

Suppose that it is required to find at least one support solution 
$\pi^{*}\in B$ such that $f(\pi^{*})$ takes a specified value. 

In a particular case, problem $Z$ is called {\it extremal} if the function 
$f(\pi^{*})$ takes an extremal value, i.e., for a maximization problem 
$f(\pi^{*})\geq f(\pi)$ and for a minimization problem $f(\pi^{*})\leq f(\pi)$, 
where $\pi\in Q$ is any support solution of problem $Z$.

\section{Auxiliary solution sets}

Denote by $W(\pi)$ the union of all feasible solutions $\pi_{l}\in Q$ of 
problem $Z$ each of which includes the solution $\pi\in Q$,
\[
W(\pi)=\bigcup_{\forall \pi_{l}\supseteq \pi} \pi_{l}.
\]

Clearly, $W(\pi)\subset R$.

The set $S(\pi)= R\setminus W(\pi)$ is called {\it adjoint} to the solution 
$\pi\in Q$. Thus, an adjoint set consists of those and only those elements of 
the work set $R$ whose union with the given feasible solution $\pi\in Q$ forms 
a subset of the set $R$ that is not a solution of problem $Z$.

\begin{theorem}
If $\pi_{1}, \pi_{2}\in Q$ and $\pi_{1}\subset \pi_{2}$, then 
$S(\pi_{1})\subset S(\pi_{2})$.
\end{theorem}

Clearly, if $\pi_{1}\subset \pi_{2}$, then $W(\pi_{2})\subset W(\pi_{1})$. 
Thus $S(\pi_{1})$ = 
$R\setminus W(\pi_{1}\subset R\setminus W(\pi_{2})=S(\pi_{2}$.$\circ$

\vspace{1pc}

The set $R(\pi) = R(\pi\cup S(\pi))$ is called the {\it residual set} for the 
solution $\pi\in Q$.

\begin{theorem}
If $\pi\in Q$ and $r\in R(\pi)\not= \oslash$, then $\pi\cup \{r\}\in Q$.
\end{theorem}

Indeed,
\[R(\pi)=R\setminus (\pi\cup S(\pi))=(R\setminus \pi)\cap (R\setminus S(\pi))=
\]
\[
=(R\setminus \pi)\cap W(\pi)=(R\cap W(\pi))\setminus \pi=W(\pi)\setminus \pi.
\]

Therefore, for $R(\pi)\not= \oslash$, the set $\pi\cup \{r\}$ is included in 
at least one feasible solution from $Q$ and is thus also a solution by property 
$1^{0}$.$\circ$

\begin{theorem}
If $\pi\in Q$ is a support solution, then $R(\pi) = \oslash$ and 
$\pi\cup S(\pi)=R$.
\end{theorem}

Let $\pi\in Q$ be a support solution of problem $Z$. Assume that 
$R(\pi)\not= \oslash$. This leads to the conclusion that the region $Q$ 
contains a solution $\pi_{1}=\pi\cup \{r\}$ $(r \in R(\pi))$, which properly 
includes the solution $\pi\in Q$. A contradiction with the definition of 
support solution.

The relationship $\pi\cup S(\pi) = R$ for every support solution $\pi\in Q$ 
follows from the definition of residual set when $R(\pi) = \oslash$.$\circ$

\section{Class of problems without lookahead}

Let $T$ be the set of all possible problems $Z$. It follows from the above 
discussion that the issue of solvability of problem $Z$ involves developing an 
algorithm (a deterministic Turing machine) that finds at least one support 
solution $\pi\in B$ of the problem for which the function $f(\pi)$ takes a 
specified value and does it in a time polynomial in the number of elements of 
the work set $R$.

In the set $T$ we identify the subclass $T_{1}$ of problems in which the set 
of support solutions contains exponentially many elements (more precisely, the 
number of elements is a function of the form $2^{p(n)}$). Any problem 
$Z\in T_{1}$ is called {\it nontrivial}. In what follows we only consider the 
set of nontrivial problems $T_{1}\subset T$ for which $Card(B) = O(2^{p(n)})$.

We say that the adjoint set $S(\pi)$ for a given solution $\pi$ is determined 
{\it effectively} if for all elements $r_{i}\in R\setminus \pi$ we can decide 
in polynomial time the truth of the predicate ``$\pi\cup \{r_{1}\}\in Q$'' or 
the predicate ``$\pi\cup \{r_{1}\}\in Q$''.

Problem $Z$ is called a {\it problem without lookahead} if for every feasible 
solution $\pi\in Q$ the adjoint set is determined effectively.

\begin{theorem}
If $Z\in T_{1}$ is a no-lookahead problem, then it is solvable by a 
nondeterministic Turing machine in polynomial time.
\end{theorem}

Indeed, by definition the size of problem $Z$ is the number of elements $n$ in 
the work set $R$, and the length of the solution, defined as the number of 
elements in some support solution $\pi\subset R$, is a linear function of $n$. 
Noting that $Z$ is a no-lookahead problem, the NTM should compute all the 
feasible solutions simultaneously. Hence it follows that every problem 
$Z\in T_{1}$ is solvable by NTM in polynomial time.$\circ$

\begin{theorem}
\label{4}
The class \np\ is identical with the class of no-lookahead problems, 
i.e., $T_{1} = \np$.
\end{theorem}

By Theorem \ref{4}, $T_{1}\subset \np$. By description, the class \np\ does not 
include inherently exponential problems. Thus, $T_{1} = \np$.$\circ$ 

\section{Conclusion}

At a first glance it would seem that the accepted interpretation of NTM 
essentially restricts the set of \np-complete problems that are considered 
when the interpretation allows solution ``guessing.'' This is not so, however. 
We know that the formulation of a problem has an essential impact on the 
possibility of solving the problem. The accepted interpretation of NTM 
operation makes it possible to reject formulations that {\it a priori} require 
exhaustive enumeration of an exponential set of support solutions.

Thus, consider the problem to find the Hamiltonian cycle in a graph. This is 
an inherently exponential problem if it is formulated so that the construction 
of each feasible solution requires ``guessing'' a correct choice, i.e., 
``advance knowledge'' of the collection of edges that forms a feasible 
solution, or belongs to at least one support solution -- a Hamiltonian cycle.

The same problem can be formulated in a different setting: find a partition of 
the graph into a minimum number of cycles and edges. If the graph is 
Hamiltonian, then the solution of this problem produces the sought cycle.

\end{document}